# Study on temperature coefficient of CdTe detector used for X-rays detection

GUO Si-Ming(郭思明)[1], WU Jin-Jie(吴金杰)[1], ZHANG Jian(张健)[1], LI Xu-Fang(李旭芳)[2], LIU Cong-Zhan(刘聪展)[2],

ZHANG Shuai(张帅)[1], LI Cheng-Ze(李承泽)[1], HUO Bin-Bin(霍彬彬)[1], LIAO Zhen-Yu(廖振宇)[1]

1. National Institute of Metrology, China, Beijing 100013, China

2. Institute of High Energy Physics, Beijing 100049, China

**Abstract**：The temperature of the working environment is a key factor in determining the properties of semiconductor detectors, and it affects the absolute accuracy and stability of the standard detector. In order to determine the temperature coefficient of CdTe detector used for X-rays detection, a precise temperature control system was designed. In this experiment, detectors and radiographic source were set inside the thermostat with temperature of 0～40℃, so that the temperature can be regulated for the test of the temperature coefficient of CdTe detector. Studies had shown that, with the increase of the temperature, the energy resolution and detection efficiency of the CdTe detector would deteriorate, and under 10℃ the detectors have better performance with the 8 keV X-rays.

## 1 Introduction

The basic principle of semiconductor detector is that the high voltage electric field is generated in the semiconductor by a thin layer of metal electrodes on the surface of semiconductor. Ionizing rays generates electron-hole pairs in the detector when the rays get in, the number of the electron-hole pairs is proportional to the energy of the incident particle. Electrons and holes move to different electrodes, and are collected by the electrodes. The electrical pulse signal is thus formed. Cadmium telluride (CdTe) is a type of compound semiconductor materials [1]. Compared with Si and Ge, CdTe has a larger atomic number and higher density (ρ =5.83g/cm3), hence it has better ability to prevent and absorb the X-rays and gamma rays, so it has higher intrinsic detection efficiency. Besides, CdTe detector also has the advantage of small size and high resolution ratio. The most remarkable characteristic of the CdTe detector is that it can work in room temperature [2].Thanks to these characteristics, CdTe detector has broad application, including medical diagnosis [3].

Performance of detector can be various at different temperatures. Temperature characteristic is a key factor that influences the absolute accuracy and stability of a standard detector. CdTe detectors can be operated at room temperature. But the temperature is changing, so the performance of detector can also be affected. [4]. Thus, it is a significant work to carry out the study on temperature characteristics of CdTe detector within 0~40°C and understand the influence of temperature on CdTe detector's detection efficiency and energy resolution.

## 2  Experimental facility

### 2.1 Temperature control system

Temperature control system is thermostat designed for the experiment. The thermostat uses the technology of compressor refrigeration, as well as stable and reliable control technology. The control system is independently researched and developed. Equipped with a high resolution ratio true color touch screen, it is user-friendly, convenient to operate and has high control accuracy. Therefore, it can provide a stable accurate temperature environment which meets the needs of the experiment. As shown in Figure 1, the thermostat temperature control system has a separated structure, the thermostat and operation screen occupy upper part, and supporting frame in lower part.  The refrigerating compressor set is separately placed behind the test chamber and connected by refrigeration pipe line. The system structure is compact and takes up small space.

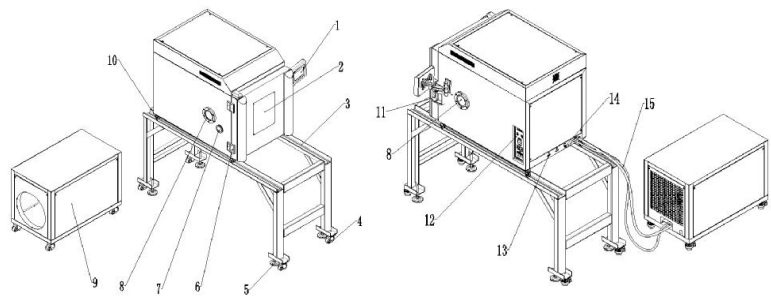

Figure 1 Diagram of the thermostat

In the figure: 1, Touch screen; 2, Viewing window; 3, Steel tank; 4, Wheel of steel tank; 5,Angular glass; 6, Hinge; 7, Wire hole; 8, Rays hole; 9, Refrigeration and heating set; 10, Wheel of thermostat; 11, Handle; 12, Control panel; 13, Power wire; 14, inflator nozzle; 15, refrigeration tubes

The thermostat has a nominal capacity for 216L, controllable temperature range -20.0~+100.0 °C, fluctuation of temperature ≤±0.5°C, uniformity of temperature ≤2°C, temperature deviation ≤±2°C, cooling rate ≥1°C/Min, (+70°C~-20°C, zero load, tested according to GB/T 5170.2-2008), warming rate ≥3°C/Min (-20°C ~+70°C, zero load, tested according to GB/T 5170.2-2008), using low-noise high efficiency scroll refrigeration compressor for refrigeration with an air-cooled engine for cooling, using high-efficient stainless steel finned tube heater for heating. The cycle air cooled machine for cooling is highly reliable noise-proof motor and stainless steel centrifugal impeller. In addition, temperature control system has security control device for electric leakage protection, phase sequence protection, incorrect operation protection, overpressure protection, overheating (overload) protection, working fluid leakage alarm, fan overheating (overload) protection, heater overheating protection, multiple temperature excess protection, program operation protection, and temperature warning. The thermostats has high stability, can precisely adjust and control the experimental environment temperature to meet experimental requirements.

**2.2 X-rays generation system**

In the experiment an X-rays apparatus is used to produce X-rays. X-rays apparatus consists of X-rays tube, power supply and control circuit. High voltage provides high pressure field makes active electrons on filament accelerated flow to anode, electron current bombarded anode target, most of the power is converted to heat, a small part of the power produce X-rays by bremsstrahlung, Figure 2 shows the principle of X-rays apparatus. In the experiment, a micro X-rays apparatus is used, its maximum tube voltage is 50kV, maximum tube current is 1.0mA, and the peak-power is 50W.

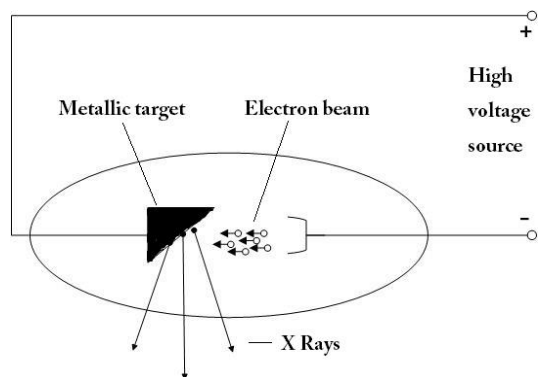

Figure 2 Principle of X-rays apparatus diagram

In X-rays tube, the cathode launches out two different X-rays produced spectrum from accelerated electrons bombardment anode target under high voltage. One is continuous spectrum, uncorrelated with target material, the other one is linear characteristic X-rays spectrum, which is uncorrelated with accelerated voltage, but correlated with target material. Characteristic X-rays spectrum is produced by inner shell electronic transitions of atoms. Inner shell electrons are ionized when high-speed electrons bombarding target atoms, a hole is produced in the inner shell, electrons in outer shell transit to the hole in inner shell where generated electromagnetic radiation spectrum, namely characteristic X-rays spectrum. Usually K, L, M, N...are used to represent principal quantum number (number of electron shells in chemistry), n=1, 2, 3, 4 ... represents energy level, When the electron at n=2 shell transits to the hole at n=1 shell, This radiation is known as the $K_\alpha$ type, When the electron at n=3 shell transits to the hole at n=1 shell, this radiation is known as the $K_\beta$ type, n=3 jump to n=2 shell called $L_\alpha$ type, n=4 jump to n=2 called $L_\beta$ type etc. It is more intuitive to understand mechanism of producing X-rays using XOP software simulation the spectrum of X-rays apparatus. Figure 3 shows X-rays spectrum of tungsten target at 20kV by XOP software simulation. In the figure, the two peaks respectively correspond to $L_\alpha$ and $L_\beta$ characteristic peak of tungsten target.

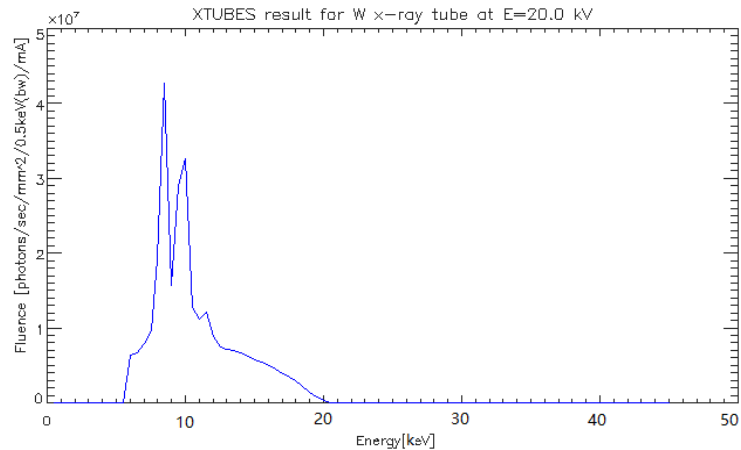

Figure 3 X-rays spectrum of tungsten target at 20kV by XOP simulation

Control software of the X-rays apparatus displays temperature of the X-rays tube in real time. Temperature of the X-rays apparatus is almost invariant when changing the temperature in the thermostat during the experiment, it always remains around 40°C. The effects on X-rays apparatus of temperature change could be omitted in the experiment.

## 3 CdTe detector temperature characteristic experiment

To study the detection efficiency of the detector should determine the absolute photon number of detector's location first. In the experiment, a standard HPGe Detector is used for absolute photon number measurement. The detection efficiency of the HPGe detector has been calculated by Monte Carlo simulation program. The photon number deposited in HPGe Crystal can be recorded in different energy interval with a defined parallel X-rays source, so the detection efficiency in each energy interval can be calculated. The results of simulation displayed that the maximum energy that can be detected of the HPGe detector is 350.0keV, the detection efficiency is high between 14.84keV and 96.23keV energy interval, can reach to more than 90%. Figure 4 shows the detection efficiency in (2-18) keV energy interval. At about 11.00keV, the detection efficiency decreased obviously because Ge elements are activated, photons cannot be documented owing to the occurrence of $K_\alpha$ and $K_\beta$ characteristic X-rays escape.

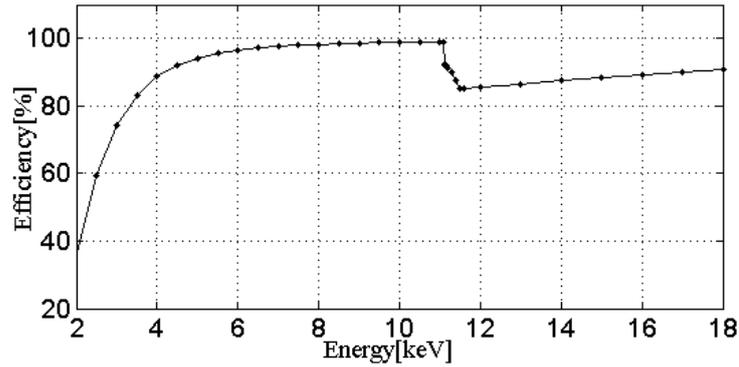
Figure 4 Detection efficiency curve of HPGe with (2-18) keV photon

Set the X-rays tube voltage at 20kV, the energy spectrum is measured with the standard HPGe detectors, as shown in Figure 5. Contrast with Figure 3, Figure 5 shows that the characteristic peak appears around 8keV with 20kV X-rays tube voltage. There are two peaks apparent in the experiment.

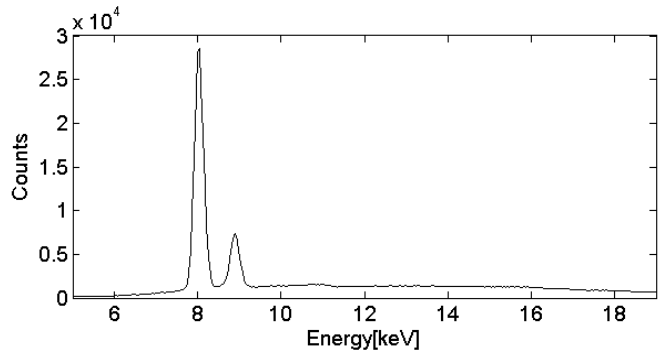
Figure 5 Measured spectrum of X-rays tube by HPGe

The temperature characteristics is studied by using a CdTe semiconductor detector, the outermost of the probe is a covering layer of beryllium window, underside are CdTe crystals, electronics system, cooling module and power system. A layer of platinum is attached to upper surface of CdTe crystal, lower surface is ceramic material for supporting, the whole detector is wrapped up by stainless steel. X-rays apparatus and CdTe detector are placed in thermostat and make sure the CdTe crystal face directly to exit portal of X-rays apparatus, experiment device schematic is shown in Figure 6. In order to guarantee the temperature of detector is same as environment temperature, counting should begin after the thermostat reached target temperature and stable for one hour.

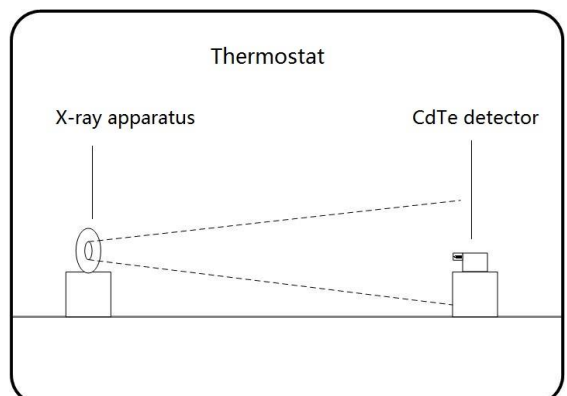
Figure 6 Diagram of CdTe temperature characteristics experiment

When adjust the temperature of thermostat from 0℃ to 40℃ with each interval of 10℃, measured spectrum of the experiment is shown in Figure 7. The bimodal spectrum shown in Figure 5 and Figure 3 appears when the temperature is 0℃. There is only one $L_\alpha$ characteristic peak when the temperature is above 10℃, meanwhile the energy resolution reduces. As the temperature increases the energy resolution becomes worse, while the detection efficiency decreases.

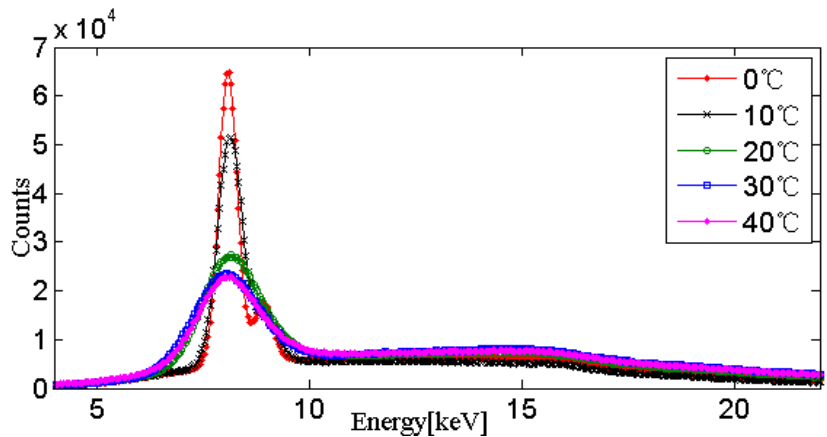

Figure 7 Spectra of X-rays tube at different temperatures by CdTe detector

The total count of the detector at the characteristic peak of $L_\alpha$ is shown in Table 1. The absolute number of photon at the energy level can be measured by standard HPGe detector, which can determine the detection efficiency of CdTe detector, and understand the effect of temperature on the detection efficiency of the detector. The counts of the peak of HPGe detection is 456,336, and counts of the peak area is 4,579,680. As the detection efficiency at the energy point is 98%, the total photon number at the peak is 465,649, and the total number of photons getting into the HPGe detector is 4,673,143. The effective detection area of this HPGe detector is 104.04 mm$^2$, and the effective detection area of the CdTe detector is 23.59 mm$^2$, so the number of number of mono-energetic photon getting into CdTe is 23.59/104.04×465,649=105,581, and the total detected photon number of the CdTe detector is 1,059,587, hence the detection efficiency of CdTe detection could be deduced, the result is shown in Table 1.

The principle of the semiconductor detector is that the photons hit the crystal atoms, and the atoms are activated to generate electron-hole pairs, collecting electron-hole pairs to generate current signal. The noise becomes larger with the increase of temperature, so the energy resolution and detection efficiency of the detector will obviously decrease. The experimental results show that the detection efficiency and energy resolution of CdTe detector will decrease with the temperature rising. At 0°C, the detection efficiency is highest and peak detection efficiency is 61.32%, the total detection efficiency reach 80.96%, energy resolution is the best, two characteristic peaks could be separated perfectly. The performance is significantly reduced with the temperature increasing. The detection efficiency and energy resolution of the detector is poor for low energy X-rays at 8keV at 20°C or higher. There is peak shift to the right with the increase of temperature which is consistent with the reference [8].

Table 1 Result statistics

| Temperature [°C] | Counts at the peak | Counts in the area | Detection efficiency at the peak [%] | Detection efficiency [%] | Energy resolution |
| --- | --- | --- | --- | --- | --- |
| 0 | 64,751 | 857,879 | 61.32 | 80.96 | 6.95% |
| 10 | 51,605 | 811,334 | 48.88 | 76.57 | 9.89% |
| 20 | 27,350 | 615,983 | 25.90 | 58.13 | 23.07% |
| 30 | 22,944 | 529,171 | 21.73 | 49.94 | 25.27% |
| 40 | 21,910 | 511,700 | 20.75 | 48.29 | 31.15% |

# 4  Conclusion

(1) Seen from the experimental result, the environment temperature has a great influence on the detection efficiency of CdTe detector. As the temperature increases, the peak count reduces accordingly. The count decreased about 66% with the temperature changing from 0°C to 40°C. The lower the temperature, the higher the detection efficiency. When the temperature is above 10℃, the detection efficiency is getting lower and lower as the temperature increases above 10°C. Within 10°C, the efficiency is acceptable and the results of the experiment are basically consistent with the theory.

(2) The energy resolution of the detector acts well below 10°C, which can remain below 10%; while the temperature rises above 20°C, the energy resolution becomes poor, and it cannot reflect the actual condition of the radioactive source.

(3) The experimental results show that this type of CdTe detector is reliable to be used in X-rays detection. The detection efficiency is very high especially under temperature below 20°C. On the basis of this experiment, we could know that it's necessary to use air conditioning to control the environment temperature when we measure the radiation with CdTe detectors. The results are especially ideal when the ambient temperature is below 10°C. In addition, according to the influence of temperature on the detection efficiency, some correction should be made to reduce the uncertainty of the study.

**References**

[1] Arlt R, Rundquist D E. Room temperature semiconductor detectors for safeguards measurements[J]. Nuclear Instruments & Methods in Physics Research, 1996, 380(1–2):455-461.

[2] Eisen Y. Current state-of-the-art industrial and research applications using room-temperature CdTe and CdZnTe solid state detectors[J]. Nuclear Instruments & Methods in Physics Research, 1996, 380(1–2):431-439.

[3] Mori I, Takayama T, Motomura N. The CdTe detector module and its imaging performance.[J]. Annals of Nuclear Medicine, 2001, 15(6):487-94.

[4] Suzuki K, Sawada T, Seto S. Temperature-Dependent Measurements of Time-of-Flight Current Waveforms in Schottky CdTe Detectors[J]. IEEE Transactions on Nuclear Science, 2013, 60(60):2840-2844.

[5] Jiang Wei. The measurement and research of personal dose equivalent for protection level X-rays. [D]. Chengdu University of Technology, 2014.

[6] Chen Cheng, Wu Jinjie, Zhou Sichun, et al. CT technology combined with MC simulation efficiency of detector calibration curve [J]. Nuclear Electronics & Detection Technology, 2015, 35(5):490-494.

[7] Matsumoto C, Takahashi T, Takizawa K, et al. Performance of a new Schottky CdTe detector for hard X-ray spectroscopy[J]. 1997, 45(3):428-432.

[8] Dusi, W, Caroli, E, Di Cocco, G, et al. A study of temperature dependence of some relevant parameters performed on a set of CdTe detectors[C]// Nuclear Science Symposium and Medical Imaging Conference. IEEE, 1994:143 - 146.